\documentclass[%
 reprint,
superscriptaddress,
 amsmath,amssymb,
 aps,
 pre
]{revtex4-1}

\usepackage{graphicx}
\usepackage{dcolumn}
\usepackage{bm}
\usepackage{comment}

\begin{document}

\preprint{APS/123-QED}

\title{Birth and Death of One-dimensional Domains in Cylindrically Confined Liquid Crystals}

\author{Madina Almukambetova}
\thanks{Equal contribution}
\affiliation{Department of Physics, Ulsan National Institute of Science and Technology, Ulsan, Republic of Korea}

\author{Arman Javadi}
\thanks{Equal contribution}
\affiliation{Department of Physics, Ulsan National Institute of Science and Technology, Ulsan, Republic of Korea}

\author{Jonghee Eun}
\affiliation{Department of Physics, Ulsan National Institute of Science and Technology, Ulsan, Republic of Korea}

\author{Juneil Jang}
\affiliation{Department of Biomedical Engineering, Ulsan National Institute of Science and Technology, Ulsan, Republic of Korea}

\author{Cheol-Min Ghim}
\affiliation{Department of Physics, Ulsan National Institute of Science and Technology, Ulsan, Republic of Korea}

\author{Joonwoo Jeong}
\email{jjeong@unist.ac.kr}
\affiliation{Department of Physics, Ulsan National Institute of Science and Technology, Ulsan, Republic of Korea}



\date{\today}

\begin{abstract}
Nematic liquid crystal (LC) is a partially ordered matter that has been a popular model system for studying a variety of topological behaviors in condensed matter. In this work, utilizing a spontaneously twisting achiral LC, we introduce a one-dimensional (1D) model system to investigate how domains and topological defects arise and annihilate, reminiscing the Kibble-Zurek mechanism. Because of the unusual elastic properties, lyotropic chromonic LCs form a double-twist structure in a cylindrical capillary with degenerate planar anchoring, exhibiting chiral symmetry breaking despite the absence of intrinsic chirality. Consequently, the domains of different handedness coexist with equal probabilities, forming the topological defects between them. We experimentally measure the domain-length distribution and its time evolution, best fitted by a three-parameter log-normal distribution. We propose that the coalescence within a train of 1D domains having the normal length distribution and randomly assigned handedness, may lead to the domains of the log-normal-like length distribution. Our cylindrically confined LC provides a practical model system to study the formation and annihilation of domains and defects in 1D.

\end{abstract}

\maketitle


\section{Introduction}

Symmetry-breaking phase transition is of great interest in various fields, ranging from cosmology to condensed matter \cite{Kibble1976, Zurek1985, Hendry1994, Weiler2008, Monaco2009, DelCampo2010, Ulm2013, Pyka2013, Deutschlander2015, Fowler2017, Bowick1994, Chuang1991, Chuang1991a}. When a system undergoes a phase transition from higher to lower symmetry phases, different regions of broken symmetries appear. If the symmetries of adjacent domains are incompatible, topological defects may form at their boundaries. Kibble \cite{Kibble1976} and Zurek \cite{Zurek1985} proposed a theoretical model to statistically understand the dynamics of defect formation and annihilation, utilizing the examples of cosmic string formation in the universe and superfluid phase transition in helium, respectively. The resulting scaling laws connect the density of defects and phase transition rate. Regardless of a system's nature, various experimental systems have validated the Kibble-Zurek mechanism and scaling laws: superfluids \cite{Hendry1994} and superconductors \cite{Monaco2009} to laser-cooled ion crystals \cite{DelCampo2010, Ulm2013, Pyka2013}, trapped atom gas \cite{Weiler2008}, and two-dimensional colloidal systems \cite{Deutschlander2015}.

Nematic liquid crystals (LCs) have been a popular system for studying the formation of domains and accompanying defects involved in phase transitions. Nematic LC with no long-range positional ordering, but the orientational order is arguably the simplest partially ordered matter and allows a facile observation of topological defects under optical microscopes. Thus, it is one of the earliest experimental systems verifying the Kibble-Zurek mechanism \cite{Chuang1991, Chuang1991a, Bowick1994}. Specifically, the formation and evolution of strings, monopoles, and textures in nematic LC films have been observed, and the Kibble-Zurek scaling laws successfully describe the relation between the defects density and the cooling rate \cite{Chuang1991,Bowick1994, Bradac2002, Fowler2017}. In the same vein, water-based lyotropic chromonic liquid crystals (LCLCs) with a broad isotropic-nematic coexistence regime have been employed to investigate the morphogenesis of nematic tactoids and defects formation by coalescence of the tactoids \cite{Kim2013,Wang2018,Golovaty2020}. 

In this work, we introduce a one-dimensional (1D) model system to investigate how domains and defects form, evolve, and annihilate. We utilize a cylindrically confined LCLC exhibiting the double-twist director configuration. A train of right- or left-handed domains is formed with topological defects between two adjacent domains of different handedness. Upon quenching from the isotropic to nematic phase, we observe nematic tactoids' nucleation, growth, and coalescence to form the train of domains with alternating handedness, reminiscing the Kibble-Zurek mechanism\cite{Kim2013} and 1D Ising model. Taking advantage of the 1D system, we investigate how the domains get coarsened by lowering the system's elastic free energy. Lastly, to justify an early-stage domain-size distribution, we propose a statistical distribution best describing our experimentally observed domain-size distribution and a theoretical model considering a stochastic merging of pre-domains.

\section{Results and Discussion}

\begin{figure*}[t]
\centering
\includegraphics{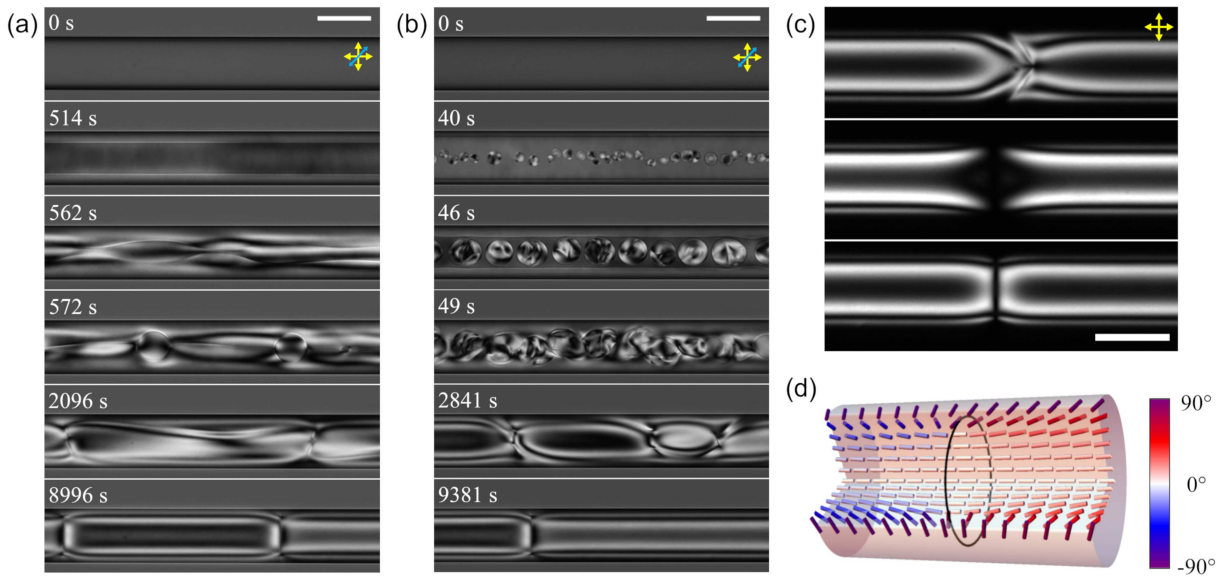}
\caption{\label{fig:formationtype}
Defects formation in cylindrically confined nematic DSCG. (a) and (b) Time-lapse polarized optical microscopy (POM) images of the 14.0\% (wt/wt) DSCG between crossed polarizers with a full-wave plate during the isotropic-to-nematic phase transition at different cooling rates: (a) 2$^{\circ}$C/min, (b) 20$^{\circ}$C/min. Yellow arrows represent pass axes of the polarizers, and a tilted blue arrow corresponds to the slow axis of the full-wave plate of 550-nm retardance, inserted before the analyzer. All scale bars are 100 $\mu$m. White texts correspond to the time elapsed after cooling. (c) POM images of three types of defects after relaxation: point (Top), domain wall (Middle), and disclination ring (Bottom). (d) Schematic diagram of the DT director configuration with the disclination ring. Rods representing the nematic director are color-coded according to the angle between the capillary axis and the director. Here red- and blue-shaded regions correspond to right- and left-handed domains, respectively, according to the right-hand rule. The black circle represents the disclination-ring defect.
}
\end{figure*}

\subsection{Formation of Double-twist Domains and Time Evolution of Their Size Distribution}

Our model system is a cylindrically confined nematic LCLC having the double-twist (DT) director configuration. Nematic LCLCs have unusual elastic properties of LCLC, such as giant elastic anisotropy and large saddle-splay modulus \cite{Zhou2012, Zhou2014, Nayani2015, Davidson2015, long2022violation}, which often give rise to chiral symmetry breaking of the director configurations under confinement despite their achiral building blocks \cite{Tortora2011, Jeong2014, Jeong2015, Davidson2015, Nayani2015, nayani2017using, fu2017spontaneous, Javadi2018, Eun2019, srinivasarao2021spontaneous}. Specifically, the nematic disodium cromoglycate (DSCG)---a representative LCLC---exhibits spontaneous chiral symmetry breaking to form the DT structure when confined in cylindrical capillaries with degenerate planar anchoring \cite{Nayani2015, Eun2019}. The directors of the DT configuration twist along the capillary radius while the directors at the capillary center lie along the capillary axis \cite{Davidson2015, Nayani2015, Eun2019}. The large saddle-splay modulus lowers the total elastic free energy by making the surface directors at the inner capillary wall align along the circumferential direction \cite{koning2014saddle,Selinger2018}. The resultant DT configurations with the right- and left-handed twist have the same energy. Thus, domains of different handednesses coexist in a capillary with the same appearance probability $P$ ($P_{\mathrm{Right}}=P_{\mathrm{Left}}=1/2$). There is a topological defect between two domains of different handedness.

\begin{figure*}[t]
\centering
\includegraphics{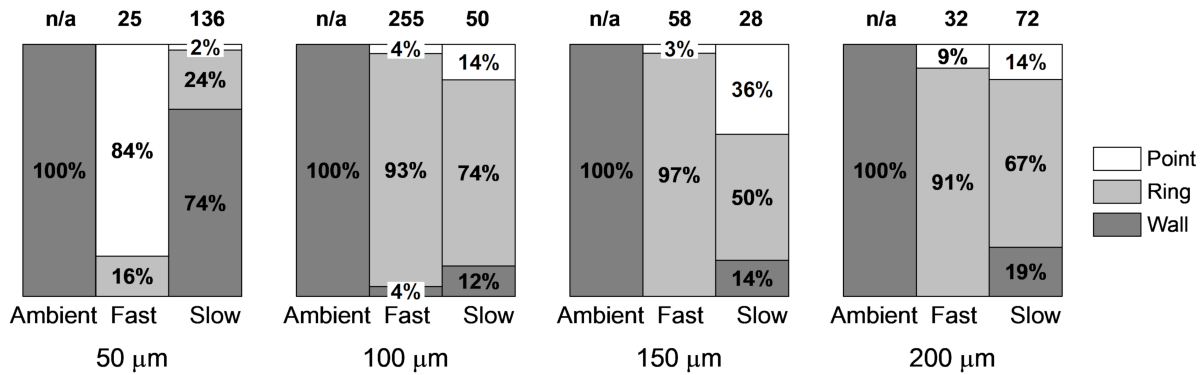}
\caption{\label{fig:defstat}
Comparison of the appearance probability of each defect type according to the cooling rate and the capillary size. We investigate the types and number of defect in capillaries of four different diameters, ranging from 50 to 200 $\mu$m. The numbers above each column indicate the total number of defects investigated; ``n/a'' means that the number is not available because we did not count them. ``Ambient'' data are from the capillaries after injecting room-temperature nematic DSCG without heating and cooling procedure, and these capillaries exhibit only ``wall'' defect. See Fig.~\ref{fig:formationtype}(c) for the types of defects. The fast and slow cooling correspond to 2 and 20$^{\circ}$C/min, respectively.
}
\end{figure*}

\begin{figure*}[t]
\centering
\includegraphics{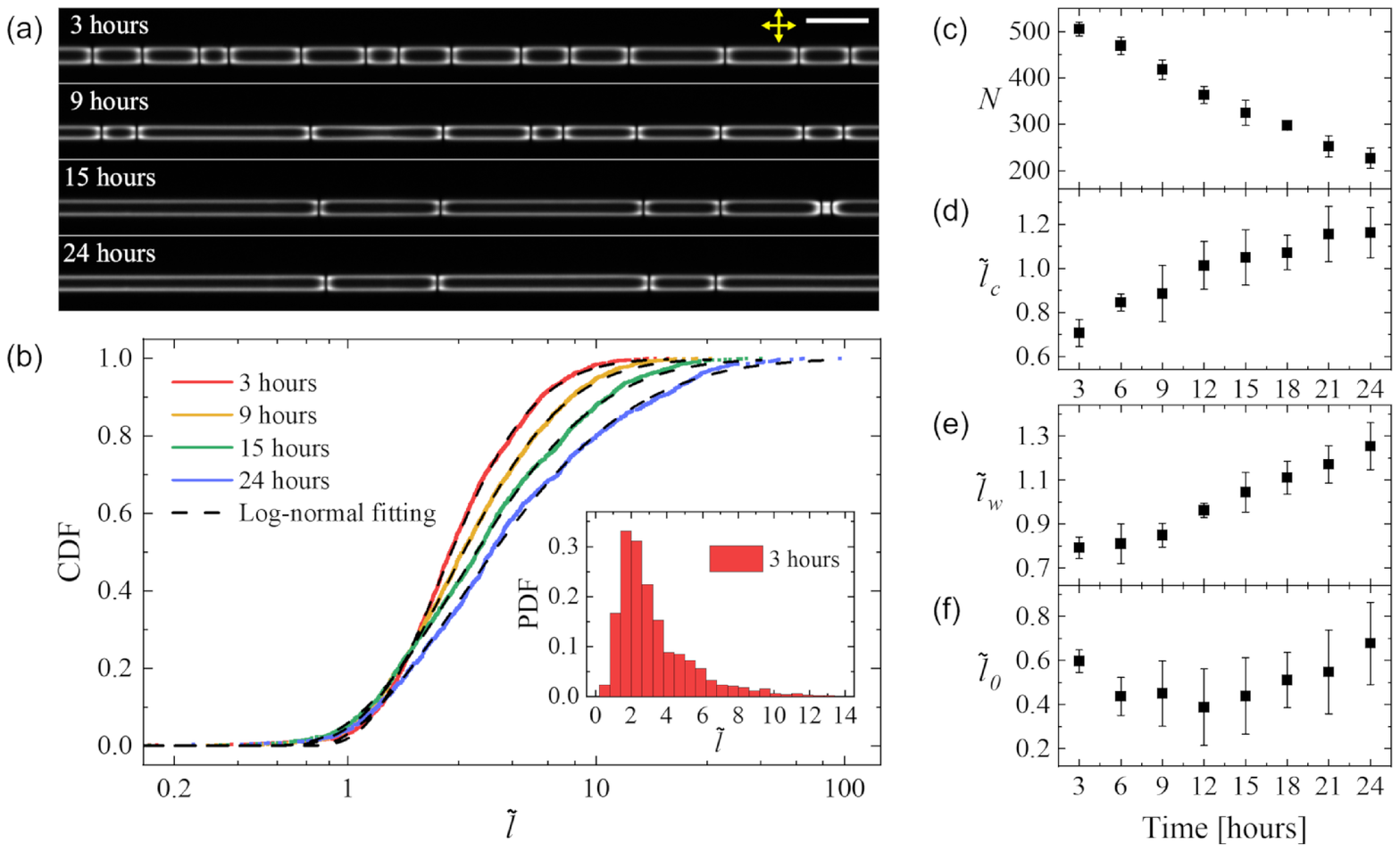}
\caption{\label{fig:tevolcdf}
Statistical characterization of the time-evolving distribution of domain sizes. The specimen is 14.0\% (wt/wt) DSCG in 100 $\mu$m-diameter cylindrical capillary after a fast cooling at the rate of 20$^{\circ}$C/min. (a) Time-lapse POM images of the DT configuration with ring defects separating the domains of opposite handedness. The scale bar is 200 $\mu$m, and polarizer axes are represented by double arrows. (b) The experimentally measured cumulative distribution function of non-dimensionalized domain sizes $\tilde{l}=l/D$ (colored solid lines). Each black dashed line corresponds to the best fit to each data with the three-parameter log-normal distribution function. An inset shows the probability distribution function of representative data at 3 hours. The data is taken from five independent experiments, and each experiment investigates six capillaries of approximately 30 mm in length at the designated time interval. (c) The total number of domains in a single experiment with six capillaries as a function of time. The symbols show the average values of the five independent experiments, and the error bar is the standard deviation. (d)-(f) The fitting parameters of the three-parameter log-normal fittings in (b): (d) mean $\tilde{l}_c$, (e) standard deviation $\tilde{l}_w$, and (f) cutoff $\tilde{l}_0$ of the logarithm of non-dimensionalized domain size $\tilde{l}$. See Eq. (\ref{eq:lncdf}) for the distribution function.
}
\end{figure*}

The domain formation process and resulting defect density strongly depend on the rate of isotropic-nematic phase transition before the nematic DT configuration forms. Fig. \ref{fig:formationtype} (a) and (b) compare two experiments of phase transition from the isotropic (45$^{\circ}$C) to nematic (21.5$^{\circ}$C) phase with different cooling rates, 2 and 20$^{\circ}$C/min, respectively. Fig. \ref{fig:formationtype}(a) shows how the nematic DT domains form at the cooling rate of 2$^{\circ}$C/min. Entering the isotropic-nematic coexistence phase upon cooling from the isotropic phase, a nematic shell grows from the capillary wall. However, as shown in Fig. \ref{fig:formationtype}(b) with the cooling rate of 20$^{\circ}$C/min, the nematic shell grows from the capillary wall, and also nematic droplets appear and grow in bulk, simultaneously \cite{Javadi2018}. The nematic texture relaxes over time to form one-dimensional (1D) DT domains eventually, while nematic domains merge and annihilate. In our experimental conditions with the glass capillary of 100-$\mu$m inner diameter, it takes approximately three hours to clearly identify defect-separated 1D domains of different twist-handedness. The linear defect density in the slow-cooling case, \textit{i.e.}, the number of defects per unit capillary length, is over ten times smaller than the density in the fast-cooling case.

This observation that slow cooling leads to larger domains than fast cooling reminisces the scaling laws in the Kibble-Zurek mechanism, but the systematic study of the Kibble-Zurek scaling law under this confinement can be pointless. As shown in Fig. \ref{fig:formationtype} (a) and (b), the nematic droplets formed only in the fast-cooling case make the texture more disordered than in the slow-cooling case, presumably resulting in the shorter domains. On the other hand, during the droplet-free slow cooling, the Rayleigh-Plateau instability of the confinement-induced cylindrical core-shell interface seems to play a major role in the determination of the domain sizes \cite{Javadi2018}. This indicates that the mechanism of the domain formation can differ by the cooling rate, thus making it difficult to expect a single governing scaling. Furthermore, because acquiring a statistically meaningful number of domains by slow cooling is impractical, we mainly study the fast-cooling case.

In DSCG's nematic DT configuration, we identify three different types of defects between neighboring domains. A previous study presents that energetics chooses the type of the defect separating the adjacent domains of opposite handednesses: either a point defect or a domain wall \cite{Davidson2015}. For instance, only point defects have been observed in another representative LCLC, Sunset Yellow (SSY) \cite{Davidson2015}. However, we find that three distinct types can appear in our 14.0\% (wt/wt) DSCG: point, domain wall, and disclination ring as shown in Fig. \ref{fig:formationtype}(c), and their appearance probability depends on factors such as capillary size and cooling rate. Fig.~\ref{fig:defstat} compares the experimentally measured appearance probability of each defect type. In this work which focuses on 100 $\mu$m-diameter capillaries filled with 14.0\% (wt/wt) DSCG quenched at the rate of 20$^{\circ}$C/min, the disclination ring shown at the bottom row of Fig. \ref{fig:formationtype}(c) is the dominant type of defect with the appearance probability greater than 95\% \cite{Eun2019}. This type of defect is not observed in SSY, and Fig. \ref{fig:formationtype}(d) sketches its proposed director configuration. Note that we sometimes observe the domain-wall defect (the middle row of Fig. \ref{fig:formationtype}(c)), which is not observed in SSY, and the DSCG's point defect (the top row of Fig. \ref{fig:formationtype}(c)) is similar but different from the one in SSY because of their broken symmetry. The defect energetics in DSCG requires further investigation in the future.

Utilizing the 1D model domains, we investigate the domain-size distribution of the fast-cooling case and its time evolution during which the domains coarsen by merging of defects as shown in Fig. \ref{fig:tevolcdf}(a). For better statistics, we simultaneously observe six different capillaries after quenching them on the same temperature-controlled slide glass, and domains in all six capillaries are analyzed as one data set. The domains from the six capillaries span $\sim$165 mm in the capillary length, and there are approximately five hundred defects after three hours of relaxation from the quenching. We observe that the domain and defect arrangements before and after the heating-and-cooling process are independent, meaning that we can erase the information via the temperature cycle and conduct independent experiments multiple times with the same specimen. Therefore, repeating the same temperature experiments, we acquire five independent data sets. Note that we study only the fast-cooling data for the size-distribution study. It is because the defect density in the slow-cooling case is too low to analyze the domain-size distribution statistically, \textit{i.e.}, ten times smaller than the fast-cooling case. 

We estimate the distribution of the dimensionless domain length $\tilde{l}$ at different times after the quenching. The length $l$ of a single domain is obtained by measuring the distance between two neighboring defects as described in Materials and Methods and is non-dimensionalized in units of the capillary diameter ($D$ = 100 $\mu$m), \textit{i.e.,} $\tilde{l}=l/D$. Fig. \ref{fig:tevolcdf}(b) shows the cumulative distribution function (CDF) of $\tilde{l}$ in all combined five data sets; The inset shows a probability distribution function (PDF) after three hours from the cooling, which is right-skewed. The coarsening shown in Fig. \ref{fig:tevolcdf}(a) is evident in the time evolution of CDF, \textit{i.e.,} the curve shifting toward the longer $\tilde{l}$ as time goes by. Fig. \ref{fig:tevolcdf}(c) shows how the number of defects, $N$, decreases during the coarsening as a function of time. The driving force of this coarsening process is the defects' elastic energy penalty; defects annihilation by merging them can lower the total energy of the system \cite{Jeong2015}. In this work, we focus on understanding the domain-size distribution from the statistical point of view, not the energetics of defect annihilation.

\subsection{Theoretical Understanding of Domain-Size Distribution}

To understand the observed domain-size distribution, we propose a theoretical model based on a stochastic merging of neighboring domains. As shown in Fig. \ref{fig:formationtype} (b), when the system enters the isotropic-nematic coexistence phase from the isotropic phase by quenching, the nematic droplets nucleate, grow, and coalesce to form the train of 1D domains shown in Fig. \ref{fig:tevolcdf}(a). We assume that the droplets have twisted bipolar director configuration, and their probability of having either right or left-handedness is $1/2$ because of the achiral nature of the LCLC \cite{Jeong2014}. If adjacent droplets as `pre-domains' have the same handedness, they merge to form a longer domain. Otherwise, they become two separate domains with a defect between them. Thus, the sizes of the pre-domains and their stochastic merging determine the number and sizes of the domains.

Although the actual merging and relaxation process seems complicated, as shown in Fig. \ref{fig:formationtype}(b), we assume that stochastic merging of 1D pre-domains having a certain length distribution determines the DT domain-size distribution at the early stage shown in the top panel of Fig. \ref{fig:tevolcdf}(a). The 1D pre-domain's non-dimensionalized size $\tilde{x}_{\mathrm{pre}} = x_{\mathrm{pre}}/D$ as a random variable has a probability distribution function (PDF) $f(\tilde{x}_{\mathrm{pre}})$; $x_{\mathrm{pre}}$ and $D$ are the length and diameter of a cylindrical pre-domain, respectively. Since the size $\tilde{x}$ as a dimensionless length is linearly additive, the dimensionless size of a `domain' by the merging of $n$ pre-domains is a random variable $\tilde{l}(n)=\sum_{i=1}^{n} \tilde{x}_{i,\mathrm{pre}}$, where the size of $i$th pre-domain is $\tilde{x}_{i,\mathrm{pre}}$ as an independent random variable. Then, the domain-size PDF $f(\tilde{l})$ is a mixture distribution $f(\tilde{l}) = \sum_{n=1}^{\infty} P(n)~f(\tilde{l}(n))$ with a PDF $f(\tilde{l}(n))$ of the random variable $\tilde{l}(n)$ and the probability $P(n)$ of having a merged domain by $n$ consecutive pre-domains. The probability $P(n)$ as a weighting factor should satisfy $\sum_{n=1}^{\infty} P(n)=1$. Given that the probability of having a left-handed pre-domain is $p$ and $1-p$ for the right-handed one, $P_{\mathrm{Left}}(n)$ is the probability of having $(n-1)$ consecutive left-handed pre-domains that follow a given left-handed pre-domain as a starting point of the domain. Additionally, the $(n-1)$ pre-domains should be followed by a right-handed pre-domain as a starting point of a new right-handed adjacent domain. This gives $P_{\mathrm{Left}}(n)=1 \times p^{n-1} \times (1-p)$, and $P(n)=(\frac{1}{2})^{n}$ for the equally probable left- and right-handedness.

\begin{figure}
\centering
\includegraphics{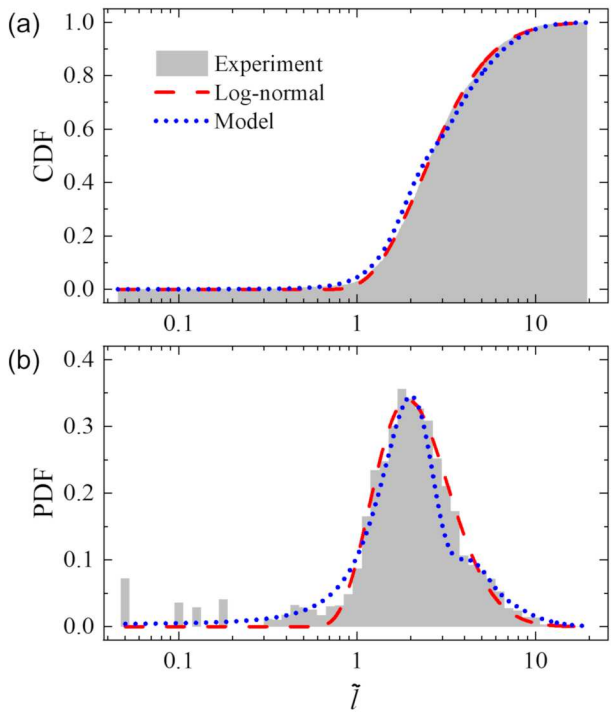}
\caption{\label{fig:compcdf}
Comparison of experimental (a) CDF and (b) PDF of non-dimensionalized domain sizes (gray solid fill) with the three-parameter log-normal fittings (red dashed line) and our theoretical model based on a  stochastic merging (blue dotted line). The experimental data is the 3-hour data in Fig. \ref{fig:tevolcdf}(b).
}
\end{figure}

We find that pre-domains having the normal size distribution can simulate approximately the experimentally measured domain-size distribution. Inspired by our experimental observation that the nucleated nematic droplets have a quite uniform size with some variation as shown in Fig. \ref{fig:formationtype}(b), we try $\tilde{x}_{\mathrm{pre}}\sim \mathcal{N}(\mu,\,\sigma^{2})$, where $\mu$ and $\sigma$ are mean and standard deviation of the normal distribution $\mathcal{N}$, respectively. Then, we have $\tilde{l}(n)=\sum_{i=1}^{n} \tilde{x}_{i,\mathrm{pre}} \sim \mathcal{N}(n \mu,\,n \sigma^{2})$ and 
\begin{equation} \label{eq:sumCDF}
F(\tilde{l})=\sum_{n=1}^{\infty} P(n)~F_{\mathcal{N}} (\tilde{l}(n)) = \sum_{n=1}^{\infty} \frac{\left[1+\textrm{erf}\left(\frac{\tilde{x}-\mu n}{\sigma \sqrt{2n}}\right)\right]}{2^{n+1}},  
\end{equation}
where $F_{\mathcal{N}}(\tilde{l}(n))$ is a cumulative distribution function (CDF) of $\mathcal{N}(n \mu,\,n \sigma^{2})$ and $F(\tilde{l})$ is the CDF of the domain-size distribution with $P(n)=(\frac{1}{2})^{n}$. In Fig. \ref{fig:compcdf}, we compare our experimental data and Eq. \ref{eq:sumCDF} to find $\mu$ and $\sigma$ numerically. With the cutoff $n_{\mathrm{max}} = 10$, $\mu = 1.693~\pm~0.004 $ and $\sigma = 0.518~\pm~0.011$ best describe the experimental data in the CDF plot shown in Fig. \ref{fig:compcdf}(a); Fig. \ref{fig:compcdf}(b) shows the fitting in the PDF plot. Instead of $\tilde{x}_{\mathrm{pre}}\sim \mathcal{N}(\mu,\,\sigma^{2})$, one may try other distributions guaranteeing positive domain size, such as Weibull or truncated normal distribution, we still find numerical similarity between the resulting CDF and experimental data (data not shown here). Note also that this comparison is only for the domain-size distribution at the early stage, i.e., the first snapshot after the relaxation into the DT configuration, because the time evolution of the distribution afterwards will be determined by the elasticity-driven interaction between the defects and resultant domain coarsening, not by the merging of pre-domains.



We find that all experimental CDFs can be approximated to the log-normal distribution's CDF (see Statistical Tests section under Materials and Methods). The dashed lines in Fig. \ref{fig:tevolcdf}(b) are the log-normal distribution fits to the time-varying experimental CDFs, and the fitting parameters are shown in in Fig. \ref{fig:tevolcdf}(d-f). The log-normal distribution is a continuous probability distribution of a random variable, whose logarithm follows the normal distribution. This log-normal nature is frequently observed in various systems such as aerosol droplets, bacteria division, avalanche processes in cells \cite{Sangal1970,Hosoda2011,Polizzi2021}. We adopt the CDF of the 3-parameter log-normal distribution

\begin{equation}
\label{eq:lncdf}
F_{\mathrm{log-normal}}(\tilde{l})=\int_{\tilde{l}_0}^{\tilde{l}}\frac{1}{\sqrt{2\pi}\tilde{l}_{\mathrm{w}} t}\exp\left(\frac{(\ln (t-\tilde{l}_0)-\tilde{l}_{\mathrm{c}})^{2}}{2\tilde{l}_{\mathrm{w}}^2}\right)dt,
\end{equation}
where $\tilde{l}_{\mathrm{c}}$ and $\tilde{l}_{\mathrm{w}}$ correspond to the mean and standard deviation of the logarithm of non-dimensionalized domain size $\tilde{l}$ after subtracting the offset $\tilde{l}_0$. As shown in Fig. \ref{fig:tevolcdf} (d-f), both $\tilde{l}_{\mathrm{c}}$ and $\tilde{l}_{\mathrm{w}}$ according to Eq. \ref{eq:lncdf} monotonically increase as time passes. $\tilde{l}_{\mathrm{c}}$ increases because the average domain size gets larger while the number of defects decreases as they annihilate by merging (See Fig. \ref{fig:tevolcdf}(a)). Note that, even at the later stage with the larger $\tilde{l}_{\mathrm{c}}$, there still exist short domains since the coarsening keeps occurring, which results in the large variance in the domain size $\tilde{l}_{\mathrm{w}}$. On the other hand, we see no consistent trend in change of $\tilde{l}_0$. Theoretically, domains shorter than $\tilde{l}_0$ should not exist. The experimental interpretation of $\tilde{l}_0$ is that domains shorter than $\tilde{l}_0$ is hardly observed because the neighboring defects at a short separation quickly annihilate by merging. This cutoff is determined by the energetics, which does not depend on time. Therefore, $\tilde{l}_0$ remains almost constant throughout the experiments.

\begin{figure*}[t]
\centering
\includegraphics{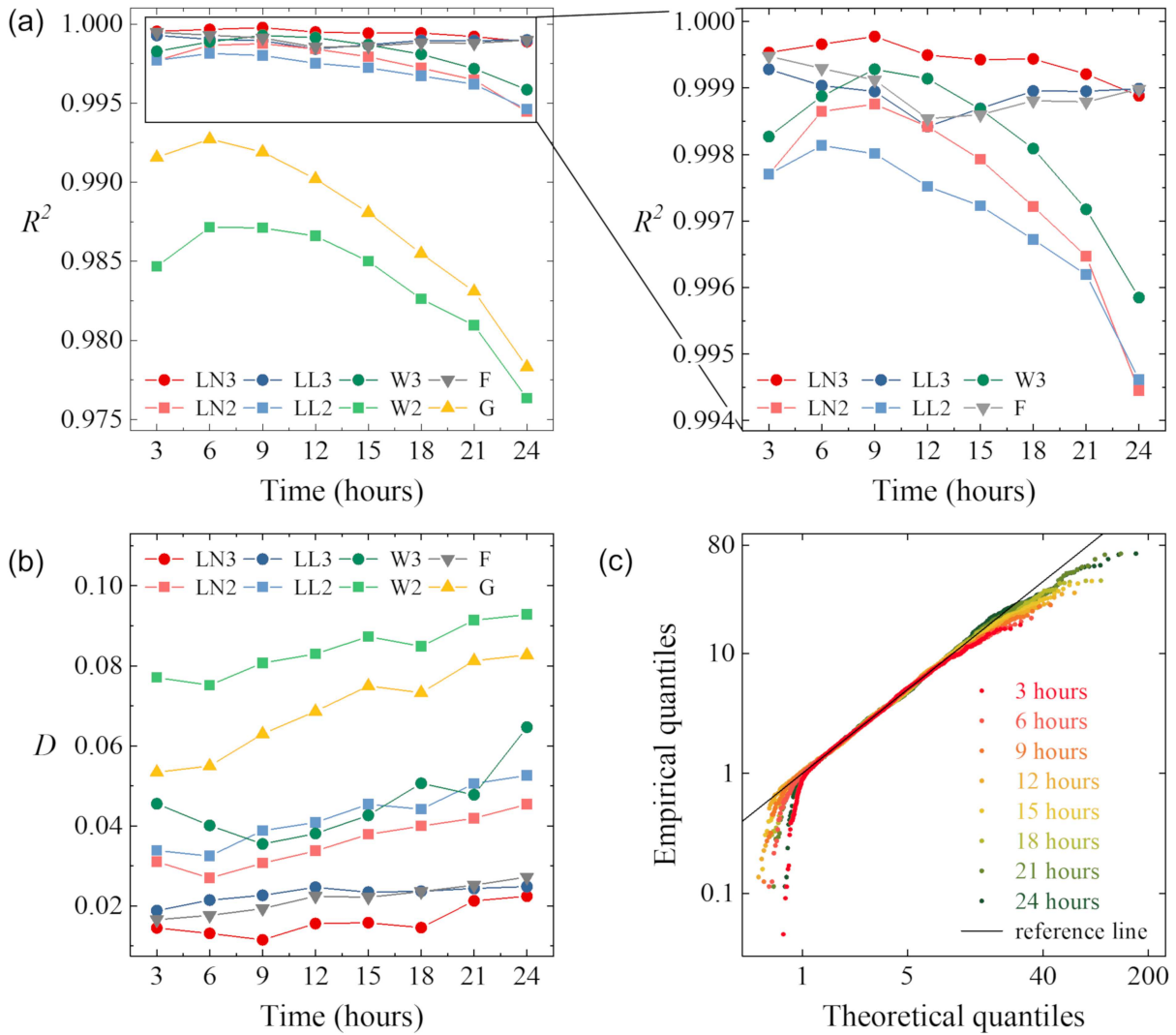}
\caption{\label{fig:COD_KS_QQ}
Two statistical tests to find the best fit to the experimental domain-size distribution shown in Fig. \ref{fig:tevolcdf}(b). Using eight right-skewed trial probability distribution functions, we evaluate (a) Coefficient of determination $R^2$ with (b) the enlarged view of the region around $R^2=1$. (b) Kolmogorov-Smirnov test statistics $KS$ of the eight trial functions. (c) Quantile-Quantile plot of the three-parameter log-normal distribution in the logarithmic scale. LN, LL, and W stand for the log-normal, log-logistic, and Weibull distributions, respectively. The following number, \textit{i.e.}, 2 or 3, indicates the number of parameters. F is for the Fr\'echet distribution with three parameters and, G for the gamma distribution with two parameters. 
}
\end{figure*}

\section{Materials and Methods}

\subsection{Sample Preparation}
Disodium cromoglycate (DSCG, purity of 95\%), acquired from Sigma-Aldrich is used without further purification. By dissolving DSCG into deionized water (18.2 M$\Omega~$cm), we prepare the 14.0\% (wt/wt) DSCG solution. We fill this solution into a cylindrical glass capillary (CV1017, Vitrocom) of 100-$\mu$m inner diameter with $\pm$10\% tolerance, at the room temperature of 22$\pm$2$^{\circ}$C with no surface treatment of the capillary. The DSCG-filled capillaries are put on a slide glass and capillaries' both ends are sealed with epoxy to minimize water evaporation during experiments. We place a coverslip on top of the capillaries and fill the space between the slide glass and coverslip with an index-matching oil (n = 1.474 at 589.3 nm, Cargille Labs).

\subsection{Optical Microscopy}
We observe the slide on a temperature controller (T95-PE120, Linkam) using a polarized optical microscope (BX53-P, Olympus). First, the sample is heated to 45.0$^{\circ}$C at the rate of 20$^{\circ}$C/min, and nematic DSCG melts into the complete isotropic phase. After incubating the sample for ten minutes at 45.0$^{\circ}$C, we decrease the sample's temperature to 21.5$^{\circ}$C while controlling the cooling rate: 20$^{\circ}$C/min for the fast cooling process and 2$^{\circ}$C/min for the slow one. Unless specified otherwise, all observations of domains are conducted at the final temperature, 21.5$^{\circ}$C, where the 14.0\% (wt/wt) DSCG solution is in the fully nematic phase.

The polarized optical microscopy (POM) images with crossed linear polarizers are taken using a CCD camera (Infinity3-6UR, Lumenera) under illumination derived from LED lamp (LED4D067, Thorlabs). As shown in the first row of Fig. \ref{fig:tevolcdf}(a), the first set of images is taken three hours after reaching 21.5$^{\circ}$C, and the time evolution is recorded typically at 3-hour intervals for 24 hours. To facilitate the measurement of the domain-size distribution in a wide field of view, we use a 4$\times$ objective lens to take and stitch multiple images covering the whole sample area; See Data Analysis for how to stitch multiple images together. For the observation of domain formation shown in Fig. \ref{fig:formationtype}(a) and \ref{fig:formationtype}(b), we use a 20$\times$ objective lens to take images at the rate of one frame per five seconds.

\subsection{Data Analysis}
To measure the domain-size distribution in a whole capillary, we stitch multiple images using ImageJ's `Grid/Collection Stitching' plugin \cite{Preibisch2009}. Then, from the intensity profile along the stitched capillary images, we identify the positions of defects and measure the distances between all neighboring defects, \textit{i.e.}, domain lengths. Specifically, as shown in the POM images under crossed polarizers such as Fig. \ref{fig:tevolcdf}(a), the defects appear as dark lines between two bright regions. A location where the second derivative of the intensity profile along the capillary is greater than a threshold value, corresponds to the defect position. Collecting these positions and distances between the neighboring defects, we estimate the domain-size distribution of $\sim$500 domains per each experiment, as shown in Fig. \ref{fig:tevolcdf}(b) and (c), and repeat the experiments five times. Note that we present the data as the cumulative distribution function with no need to set the bin size. 

\subsection{Statistical Tests}

Testing eight right-skewed probability distributions to fit our experimentally measured distributions, we find that the log-normal distribution with three parameters best describes the experimental data. As shown in Fig. \ref{fig:compcdf}(b), the experimental probability distribution function(PDF) is right-skewed. We choose eight distributions of a similar shape to fit our experimental data from 3 to 24 hours: log-normal, log-logistic, Weibull, Fr\'echet, and gamma distributions. For log-normal, log-logistic, and Weibull distributions, we test both the typical two-parameter distribution and the three-parameter distribution with an additional offset parameter. Avoiding the binning issue, we use CDFs for fitting and test the goodness of the fits by two methods: the coefficient of determination, $R^2$ (Fig. \ref{fig:COD_KS_QQ}(a)), and the Kolmogorov-Smirnov test (Fig. \ref{fig:COD_KS_QQ}(b)) \cite{Young1977,Pobocikova2017}.

Both tests suggest that the three-parameter log-normal distribution (LN3) best describes our experimental data. We calculate the coefficient of determination $R^2 = 1- \frac{\mathrm{SS}_{\mathrm{R}}}{\mathrm{SS}_{\mathrm{T}}}$ where the residual sum of squares $\mathrm{SS}_{\mathrm{R}} = \sum_{i=1}^{N}\left[ F(\tilde{l}_i)-F_\mathcal{D}(\tilde{l}_i) \right]^2$ and total sum of squares $\mathrm{SS}_{\mathrm{T}} = \sum_{i=1}^{N}\left[ F(\tilde{l}_i)-\bar{F}(\tilde{l})\right]^2$. Here, the experimental CDF is $F(\tilde{l}_i)$ when $\tilde{l}_i$ is the $i$th nondimensionalized domain length, with its mean $\bar{F}(\tilde{l})=\frac{1}{N}\sum_{i=1}^{N}F(\tilde{l}_i)$. The CDF of a certain model distribution $\mathcal{D}$ is $F_{\mathcal{D}}(\tilde{l})$. Fig. \ref{fig:COD_KS_QQ}(a) shows $R^2$ of experimental CDFs at different times to eight trial distribution functions. Except for the two-parameter Weibull and gamma distributions, the values of $R^2$ are close to one; $R^2=1$ means the distribution model matches the data. The LN3 gives the highest $R^2$.

The Kolmogorov-Smirnov test determines the distribution best describing the experimental data based on the largest absolute difference between the experimental and model CDFs, \textit{i.e.}, ${KS} = \sup_{\tilde{l}} \left| F_{\mathcal{D}}(\tilde{l})-F(\tilde{l}) \right|$. Here $\sup_{\tilde{l}}$ denotes the supremum of the set of absolute CDF differences; the supremum stands for the value greater or equal to all the other values in a set. If the test statistic ${KS}$ is smaller than a critical value, the null hypothesis that the data comes from the particular distribution is not rejected. In our test, the critical value for the test statistic is not determined conventionally since the parameters of the fitting distributions are estimated from the data itself. Therefore, we consider the distribution with the smallest ${KS}$ to be the best model. Fig. \ref{fig:COD_KS_QQ}(b) shows ${KS}$ for all trial distributions as a function of time, which reconfirms that the LN3 fits our data best. The Quantile-Quantile (Q-Q) plots in Fig. \ref{fig:COD_KS_QQ}(c) compare the experimental and LN3 CDFs. Except for the regions of very short and long domains where the log-normal distribution expects more domains, the experimental data points lie mostly on a 45-degree reference line, indicating a fair agreement between the experimental distribution and LN3.

\section{Conclusion}

We study how nematic DT domains in a cylindrical capillary form and grow by coarsening, focusing on their size distribution. Quenching from the isotropic phase leads to the domains of different handedness separated by a singular defect. The domains get coarsened by the annihilation of defects by merging and exhibit a time-evolving length distribution. We propose a model describing how stochastic merging within a train of nematic droplets as pre-domains with normally distributed sizes and randomly assigned handedness, determines the initial size distribution after the isotropic-nematic phase transition. Our statistical tests propose that the three-parameter log-normal distribution best approximates our experimentally measured size distributions of the initial and coarsened domains. Our results manifest that the proposed 1D system allowing facile optical observation and size characterization, provides a useful model platform to study symmetry-breaking phase transition in condensed matter, accompanying domains and topological defects formation.

Looking to future works, we believe understanding the energetics of DT domains and their defects, would deepen our understanding of how the domain-size distribution is determined and evolve as time goes by. Specifically, the topological defects where the handedness of the ground-state DT domain changes have an energy penalty, thus exhibiting effective attraction with each other to lower the elastic free energy. Investigating the energy landscape may give us hints at why the coarsening dynamics maintains the log-normal size distribution over time evolution. It would also be intriguing to tune the energy landscape by adding chiral dopants or applying external fields and see how the coarsening dynamics would change \cite{Eun2019,Dietrich2017}. Then, this 1D model system with known and tunable energetics may shed light on domain formation and coarsening dynamics in complicated systems of higher dimensions.

\end{document}